\documentclass{article}

\usepackage{arxiv}

\usepackage[utf8]{inputenc} 
\usepackage[T1]{fontenc}    
\usepackage{lmodern}        
\usepackage{hyperref}       
\usepackage{url}            
\usepackage{booktabs}       
\usepackage{amsfonts}       
\usepackage{nicefrac}       
\usepackage{microtype}      
\usepackage{graphicx}

\title{Data Science Transfer Pathways from Associate's to Bachelor's
Programs}

\author{
    Benjamin S. Baumer
    \thanks{More information about the Massachusetts Data Science
Pathways project can be found at
\url{https://dsc-wav.github.io/ma-ds-pathways}}
   \\
    Statistical \& Data Sciences \\
    Smith College \\
  Northampton, MA 01063 \\
  \texttt{\href{mailto:bbaumer@smith.edu}{\nolinkurl{bbaumer@smith.edu}}} \\
   \And
    Nicholas J. Horton
   \\
    Mathematics \& Statistics \\
    Amherst College \\
  Amherst, MA 01002 \\
  \texttt{\href{mailto:nhorton@amherst.edu}{\nolinkurl{nhorton@amherst.edu}}} \\
  }

\providecommand{\tightlist}{%
  \setlength{\itemsep}{0pt}\setlength{\parskip}{0pt}}

\newlength{\cslhangindent}
\newlength{\csllabelwidth}
\newlength{\cslentryspacingunit} 
\newenvironment{cslreferences}%
  {\setlength{\parindent}{0pt}%
  \everypar{\setlength{\hangindent}{\cslhangindent}}\ignorespaces}%
  {\par}
 {
  \setlength{\parindent}{0pt}
  \ifodd #1
  \let\oldpar\par
  \def\par{\hangindent=\cslhangindent\oldpar}
  \fi
  \setlength{\parskip}{#2\cslentryspacingunit}
 }%
 {}
\usepackage{calc}

\usepackage{xcolor,soul}
\usepackage{xspace}
\usepackage{booktabs}
\usepackage{longtable}
\usepackage{array}
\usepackage{multirow}
\usepackage{wrapfig}
\usepackage{float}
\usepackage{colortbl}
\usepackage{pdflscape}
\usepackage{tabu}
\usepackage{threeparttable}
\usepackage{threeparttablex}
\usepackage[normalem]{ulem}
\usepackage{makecell}
\usepackage{xcolor}
\begin{document}
\maketitle

\begin{abstract}
A substantial fraction of students who complete their college education
at a public university in the United States begin their journey at one
of the 935 public two-year colleges. While the number of four-year
colleges offering bachelor's degrees in data science continues to
increase, data science instruction at many two-year colleges lags
behind. A major impediment is the relative paucity of introductory data
science courses that serve multiple student audiences and can easily
transfer. In addition, the lack of pre-defined transfer pathways (or
articulation agreements) for data science creates a growing disconnect
that leaves students who want to study data science at a disadvantage.
We describe opportunities and barriers to data science transfer
pathways. Five points of curricular friction merit attention: 1) a first
course in data science, 2) a second course in data science, 3) a course
in scientific computing, data science workflow, and/or reproducible
computing, 4) lab sciences, and 5) navigating communication, ethics, and
application domain requirements in the context of general education and
liberal arts course mappings. We catalog existing transfer pathways,
efforts to align curricula across institutions, obstacles to overcome
with minimally-disruptive solutions, and approaches to foster these
pathways. Improvements in these areas are critically important to ensure
that a broad and diverse set of students are able to engage and succeed
in undergraduate data science programs.
\end{abstract}


\newcommand{\R}{\textsf{R}\xspace}
\newcommand{\pkg}[1]{\texttt{#1}}
\newcommand{\nick}[1]{\sethlcolor{pink}\hl{[NH]: #1}}
\newcommand{\ben}[1]{\sethlcolor{cyan}\hl{[BB]: #1}} 
\newcommand{\XX}[1]{\sethlcolor{red}\hl{[XX]: #1}}

Accepted for publication in the \emph{Harvard Data Science Review}

\hypertarget{sec:intro}{%
\section{Introduction}\label{sec:intro}}

Two-year colleges (historically known as community colleges or junior
colleges) play a critical role in higher education in the United States.

As of 2020, these 935 public institutions enroll more than 4.7 million
students (Duffin 2021). Two-year colleges provide associate's degrees
that lead directly to employment, as well as options to transfer to
bachelor's programs. In our home state of Massachusetts, the enrollment
at the 15 public two-year colleges is comparable to that of the
undergraduate enrollment of the University of Massachusetts system.
While average tuition varies considerably by state, two-year colleges
are the most effective and affordable option for many students.

Blumenstyk (2021) describes two-year colleges as:

\begin{quote}
``the keystone for the nation's plan to help more people earn a
postsecondary credential.''
\end{quote}

By all accounts, job prospects in data science are excellent, due to
high salaries, expansive job growth, and comfortable working conditions.
According to Glassdoor, data scientist is
\href{https://www.glassdoor.com/List/Best-Jobs-in-America-LST_KQ0,20.htm}{the
\#3 job in America for 2022}, and has ranked among the top three every
year since 2016. The US Bureau of Labor Statistics reports
\href{https://www.bls.gov/oes/current/oes152098.htm}{a mean annual wage
of \$103,930} for data scientists, and estimates that jobs will grow
\href{https://www.bls.gov/ooh/computer-and-information-technology/computer-and-information-research-scientists.htm}{22\%
for Computer and Information Research Scientists} and
\href{https://www.bls.gov/ooh/math/mathematicians-and-statisticians.htm}{33\%
percent for Mathematicians and Statisticians} over the next ten years. A
number of companies have reported that they can't find sufficient
skilled candidates for these positions
(\url{http://oceansofdata.org/projects/mentoring-new-data-pathways-community-colleges}).
Due to the nature of the work, data scientists have adapted smoothly to
working remotely, an increasingly relevant factor that should only
improve employment prospects. The high probability of financial success
for graduates in data science stands in stark contrast to the
increasingly dim prospects for many master's students in other fields.
Korn and Fuller (2021) conclude that 38\% of master's programs at
top-tier private universities in the U.S. don't deliver on the promise
of earnings that exceed debt incurred to pay for tuition.

Providing equitable access to these desirable jobs is a challenge that
is symptomatic of larger issues of class and income inequality in the
United States. Several national reports (e.g., Rawlings-Goss et al.
(2018), National Academies of Science, Engineering, and Medicine (2018),
and National Academy of Engineering and National Academies of Sciences,
Engineering, and Medicine (2016)) recognize this challenge and call for
tighter partnerships between two- and four-year colleges. If the field
of data science is serious about diversifying its workforce, then there
must be paths to high-paying jobs in data science that begin at two-year
colleges, which enroll a much larger fraction of historically
under-served students than four-year colleges.

The ongoing National Science Foundation's (NSF)
\href{https://www.nsf.gov/pubs/2021/nsf21523/nsf21523.htm}{Data Science
Corps (DSC) program} focuses on creative approaches to developing a
competitive and diverse workforce in data science. Through our roles as
leaders of the NSF-funded
\href{https://dsc-wav.github.io/www/index.html}{DSC-WAV (Wrangle,
Analyze, Visualize)} program we have had the opportunity to engage in
data science projects with community organizations and to work with
partners at several two-year colleges to foster new courses and
programs. This work included organizing a Symposium on Data Science at
Massachusetts Two-Year Colleges for academic leaders on June 13, 2022
and faculty development workshops in 2021 and 2022.

\hypertarget{our-contribution}{%
\subsection{Our contribution}\label{our-contribution}}

The purpose of this paper is to help foster connections between two- and
four-year institutions that will lead to more transparent and flexible
pathways to bachelor's degrees in data science. While we use
Massachusetts as our primary example, we believe that that the insights
and approaches we suggest may be useful to other states. We focus
exclusively on data science, including cognate disciplines of
mathematics, computer science, and statistics only as they relate to
data science.

We begin by briefly surveying the landscape of data science in higher
education nationally (Section \ref{sec:related}). The lack of existing
transfer pathways make a bachelor's degree in data science burdensome
for a two-year college student to achieve without significant---and
probably unreasonable---foresight and perseverance through
administrative and bureaucratic obstacles. We use two hypothetical
community college students named Alice and Bob to illustrate how these
obstacles impede student progress. In Section \ref{sec:pathways}, we use
the bachelor's program in data science at UMass-Dartmouth (which we see
as representative of a curricular consensus in data science) as an
example, analyze potential transfer pathways, and identify five points
of friction. Our analysis leads directly to recommendations that could
provide explicit pathways in data science with relatively few new
courses and modest impact on existing programs (Section
\ref{sec:proposal}). We conclude with final thoughts in Section
\ref{sec:conclusion}.

\hypertarget{sec:related}{%
\section{Background and related work}\label{sec:related}}

\hypertarget{data-science-programs-in-higher-education}{%
\subsection{Data science programs in higher
education}\label{data-science-programs-in-higher-education}}

Since Cleveland (2001)'s action plan for data science, the field has
continued to blossom within academia. Academic data science can be
aspirationally described using a pyramid, with doctoral degrees rare but
important for leadership and research in the field. Master's degrees are
the next level, with larger numbers and considerable job opportunities.
For established disciplines, bachelor's programs (offered at four-year
colleges) and associate's programs (offered at two-year colleges), make
up the third and fourth levels of the pyramid, with larger and larger
numbers of students obtaining these degrees. Jobs are available at each
level, with the potential for interested students to pursue more
advanced degrees in order to deepen skills and expand their work
opportunities. However, workforce opportunities remain opaque to too
many students.

As an emerging discipline, data science has not yet matured to that
extent, with master's programs leading the way, bachelor's programs on
the rise, and associate's program lagging behind.

Several doctoral programs in data science now exist in the United States
(National Academies of Sciences, Engineering, and Medicine 2020) and
their graduates are now beginning academic and workforce careers.

Far more common are master's programs in data science and data
analytics, which are offered by many universities (both online and
in-person). Nationally, the growth in the number of master's degrees
granted in analytics and data science
\href{https://analytics.ncsu.edu/?page_id=4184}{is dramatic}, with more
than 45,000 degrees reported in 2020 by the Institute of Advanced
Analytics.

While the study of data science at the graduate level continues to
evolve, its footprint is already substantial. The growth of these
programs makes it possible for students at the undergraduate level to
more easily identify future programs of graduate study. What
undergraduate majors should best prepare a student for graduate study in
data science? Computer science, statistics, and mathematics are the
closest cognate disciplines, and while statistics is not always
available as an undergraduate major, it is taught everywhere and can be
folded into either a computer science or mathematics major, both of
which are available at virtually any institution.

Historically rarer (but increasingly less so) are bachelor's degrees in
data science and related fields (e.g., data analytics). These programs
make up the next level of the pyramid, with larger numbers of students
potentially entering the workforce (National Academies of Science,
Engineering, and Medicine 2018). The options---which are certain to grow
in the coming years---already provide two-year college students who are
interested in data science with visible future programs of study.

Gould et al. (2018) identifies six associate's degree programs in three
states, including New Hampshire, Pennsylvania, and Minnesota. A number
of exemplary associate's data science programs have been established in
recent years (Amstat News 2022). Many others have been created across
the nation.\footnote{The
  \href{https://academicdatascience.org/leaders/dsinstitutionupdates}{Academic
  Data Science Alliance Data Science} and the
  \href{https://amatyc.org/page/DataResources}{American Mathematical
  Association of Two-Year Colleges (AMATYC)} have created listings that
  are likely incomplete. Unfortunately, no comprehensive census of
  programs is readily available.}

\hypertarget{key-concepts-in-two-year-college-education}{%
\subsection{Key concepts in two-year college
education}\label{key-concepts-in-two-year-college-education}}

Two-year college students typically pursue associate's degrees that come
in two flavors: terminal or transfer. Many associate's degrees are
\emph{terminal} (often called associate's-to-workforce), in that they
are designed to prepare students for employment directly upon
completion. Other associate's degrees are designed to prepare students
for a smooth transfer to a four-year institution (and even a specific
bachelor's degree program at that institution) upon completion. For
example, Springfield Technical Community College offers multiple degrees
in computer science. The
\href{https://www.stcc.edu/explore/programs/cset.as/}{Computer Systems
Engineering Tech program} prepares students for various systems
administration jobs after two years of study. Conversely, the
\href{https://www.stcc.edu/explore/programs/csci.as/}{Computer Science
Transfer program} prepares students to \emph{transfer} to a bachelor's
program in computer science, with most students presumably intending to
transfer to one of the UMass campuses.

We use the term \emph{pathway} to describe a route that a two-year
college student could take to obtain a bachelor's degree. Associate's
degrees designed for transfer, as described above, are the most
well-trod starting places for such pathways. But even with an
associate's degree for transfer in hand, pathways are not always
obvious. Many states, including Massachusetts and California, have
highly visible public websites that map transfer pathways from two-year
colleges to public four-year colleges. However, not all of these
many-to-many possible pathways are mapped. For example, Bunker Hill
Community College offers a
\href{https://catalog.bhcc.edu/preview_program.php?poid=90}{Computer
Science Transfer associate's degree}, but there is no corresponding
mapping to any of the UMass campuses in the system (see Section
\ref{sec:pathways}).

Further complicating matters are \emph{articulation agreements}, which
provide an explicit transfer pathway between one specific associate's
degree program and one specific bachelor's program. These agreements may
be negotiated between public or private four-year institutions. While
these one-to-one articulation agreements are helpful, they are not as
visible as the many-to-many mapped pathways.

\hypertarget{outcomes-for-undergraduates}{%
\subsection{Outcomes for
undergraduates}\label{outcomes-for-undergraduates}}

The choice of which flavor of associate's degree to pursue has
consequences for the two-year college student. Many workforce roles for
data scientists exist at the bachelor's level (De Veaux et al. 2017;
National Academies of Science, Engineering, and Medicine 2018), and the
number is growing (Gould et al. 2018).

For those who choose further study, the bachelor's-to-master's
transition is characterized by flexibility and adaptation, because
graduate schools know that they will receive applications from students
who attended a wide variety of undergraduate schools, and who studied
highly variable subjects therein. Moreover, bachelor's programs
typically involve at least 120 credit hours of study, which often
provides ample flexibility for a student to deviate from any pre-defined
curricular path. From our own experiences, we know that it is not
uncommon for a traditional bachelor's student to major in say,
economics, only to then decide before their senior year that they want
to pursue a master's degree in data science, load up on statistics and
computer science courses in their senior year, and still put together a
competitive graduate school application.

It is important to remember that dramatically less flexibility is
available for the associate's-to-bachelor's transition, since for
two-year college students, every credit counts. We recognize that for
most two-year college students, any credit that doesn't count towards
their associate's degree program or their pre-defined transfer pathway
may be considered a ``waste'' of both time and money. California has
been a leader in fostering smoother articulation of courses between
two-year and four-year institutions (see \url{https://assist.org}). But
while the California system provides a clear solution for existing
pathways, the larger difficulties with transfer pathways are
longstanding (Blumenstyk 2021). In Massachusetts, although most students
who enroll in two-year college program after high school intend to
transfer to a bachelor's degree program, relatively few actually do so
(Murnane et al. 2022).

Longer-term, alternative options, including associate's-to-workforce
programs (Rawlings-Goss et al. 2018; Gould et al. 2018) are desirable
but outside the scope of this paper. Associate's programs in
cybersecurity, information technology, and web development---designed as
terminal degrees---have proven effective in workforce development and
the same potential exists for data science.\footnote{In contrast to the
  45,000 master's graduates in data science referenced earlier,
  according to Statista (Duffin 2021), there were more than a million
  associate's degree recipients in the United States during the
  2018-2019 academic year.We believe that associate's degree candidates
  represent an untapped data resource.}

\hypertarget{data-science-curricula}{%
\subsection{Data science curricula}\label{data-science-curricula}}

Undergraduate curricula in data science are now beginning to coalesce.
De Veaux et al. (2017) provide curriculum guidelines for undergraduate
majors in data science that are endorsed by the American Statistical
Association. The ``Data Science for Undergraduates: Opportunities and
Options'' consensus study (National Academies of Science, Engineering,
and Medicine 2018) provided a number of recommendations and findings
relevant to undergraduate data science programs and outlined key aspects
of \emph{data acumen}. The Association for Computing Machinery (ACM)
Data Science Task Force enumerated computing competencies for
undergraduate data science curricula (Danyluk et al. 2021), and syllabi
from example courses. Gould et al. (2018) provides curricular guidelines
for two-year college programs in data science. Comprehensive textbooks
(Wickham and Grolemund 2016; Baumer, Kaplan, and Horton 2021) and course
materials (Çetinkaya-Rundel 2020) support the teaching of a variety of
different introductory data science courses. Donoho (2017) ruminates on
the nature of data science as a standalone scientific discipline.

In 2019, the National Center for Education Statistics unveiled a new
series of
\href{https://nces.ed.gov/ipeds/cipcode/default.aspx?y=56}{Classification
of Instructional Programs} (CIP) codes for data science
(\href{https://nces.ed.gov/ipeds/cipcode/cipdetail.aspx?y=56\&cipid=92953}{30.70}).
These new codes allow the federal government to track the growth of
programs in data science and should result in an improved ability to
quantify how many students are studying data science.\footnote{Until
  recently, the new CIP codes were not classified as STEM disciplines,
  which had negative implications for the immigration status of
  international students. Efforts by the Academic Data Science Alliance
  and others led to reclassification of the data science CIP code.}

In what might be an important stamp of legitimacy, ABET (Accreditation
Board for Engineering and Technology) has begun
\href{https://www.abet.org/diving-into-data-science/}{accrediting its
first undergraduate data science programs}, with plans to expand to the
graduate and associate's levels.

\hypertarget{the-dsc-wav-project}{%
\subsection{The DSC-WAV project}\label{the-dsc-wav-project}}

While our interest in data science education is longstanding and
well-documented, our specific interest in two-year college pathways in
data science is motivated by our involvement in the
\href{https://dsc-wav.github.io/www/index.html}{Data Science Corps
(DSC): Wrangle, Analyze, Visualize (WAV) project} (Horton et al. 2021).
The first arm of the NSF-funded program links teams of undergraduate
students (often data science majors) at the Five Colleges (Amherst,
Hampshire, Mount Holyoke, and Smith Colleges plus the University of
Massachusetts-Amherst) with local, community-based organizations in the
service of a real-world data science problem. Legacy et al. (2022)
details how this program supports the growth of DSC-WAV student
participants.

As the Data Science Corps is a workforce development initiative, the
DSC-WAV project has an additional goal of growing and diversifying the
data science workforce. In this fast-growing segment of the economy,
\href{https://www.usnews.com/education/best-colleges/articles/why-more-colleges-are-offering-data-science-programs}{highly-satisfying,
high-paying jobs are plentiful}. After several years of working closely
with our partners at Holyoke, Greenfield, and Springfield Technical
Community Colleges on a variety of curricular- and student-focused
issues, our attention is now centered on the pathway predicament. We
believe that while the obstacles to transfer pathways in data science
are formidable, we can overcome them with relatively non-disruptive
changes. Our current focus is to help identify impediments to creating
flexible and transparent transfer agreements between two-year colleges
and public universities in Massachusetts.

\hypertarget{sec:pathways}{%
\section{Pathways to data science}\label{sec:pathways}}

Explicit transfer pathways are important. The ``Data Science for
Undergraduates'' consensus report (National Academies of Science,
Engineering, and Medicine 2018) recommends that: ``Academic institutions
should provide and evolve a range of educational pathways to prepare
students for an array of data science roles in the workplace'' and that
``Four-year and two-year institutions should establish a forum for
dialogue across institutions on all aspects of data science education,
training, and workforce development.''

Transfer pathways for mature disciplines are well established
nationally. For example, the California Assist system
(\url{https://assist.org/}) and the Massachusetts MassTransfer A2B
Degree maps (\url{https://www.mass.edu/masstransfer/a2b}) provide
students in those states with easy-to-navigate, public listings of
transfer pathways within their respective public higher education
systems. These websites allow any student to select a two-year college,
a public university, and an intended bachelor's degree field. The system
will then return a list of mapped pathways which have been pre-approved
by the state's Board of Higher Education for transfer. While less formal
or one-to-one articulation agreements between individual programs may
permit direct transfer from a two-year college to a four-year program,
these public mapping systems are the best mechanisms for broadcasting
important signals to prospective students that their academic plan is
sound. Unfortunately, few pathways in data science exist.

A handful of articulation agreements between two-year and public
four-year colleges do exist. In New Hampshire, the
\href{https://manchester.unh.edu/admissions/transfer-students/transferring-community-college}{Public
Pathways program} at the University of New Hampshire at Manchester
provides explicit transfer pathways from
\href{https://manchester.unh.edu/sites/default/files/media/2019/05/gbcc_analytics_pathways.pdf}{Great
Bay},
\href{https://manchester.unh.edu/sites/default/files/media/2019/05/mcc-analytics-pathways.pdf}{Manchester},
and
\href{https://manchester.unh.edu/sites/default/files/media/2019/05/ncc-analytics-pathways.pdf}{Nashua}
Community Colleges to their bachelor's degree in
\href{https://manchester.unh.edu/program/bs/analytics-data-science-major-data-science-option}{Analytics
and Data Science}. While these opportunities are advertised by NH
Transfer, they all point to UNH-Manchester (in contrast to the multi-way
systems in California and Massachusetts).

Fortunately, as a result of existing mapped pathways for data science
adjacent disciplines like mathematics and computer science, many courses
relevant to data science exist and are easy to transfer. This includes
mathematics and statistics courses (e.g., statistics, calculus, linear
algebra, and discrete math) along with computer science courses (e.g.,
computer science I and II, data structures and algorithms). These
existing course mappings provide a solid foundation for a transfer
pathway in data science---but they are not enough.

Our analysis of the gaps reveals \textbf{five points of curricular
friction}:

\begin{enumerate}
\def\labelenumi{\arabic{enumi}.}
\tightlist
\item
  A first course in data science (Data Science I)
\item
  A second course in data science (Data Science II)
\item
  A course in scientific computing, data science workflow, and/or
  reproducible computing
\item
  Lab sciences
\item
  Navigating communication, ethics, and application domain requirements
  in the context of general education and liberal arts course mappings
\end{enumerate}

Our analysis comes with recommendations for solutions. Some of these
points of friction are best solved through new course offerings at the
two-year college level. Others are likely negotiable using existing
courses through careful planning, advising, and sequencing. Other
sources of friction, such as institutional inertia, faculty development
and retention, and technology, are no less real, but are not our focus
in this paper.

In the remainder of this Section, we use real-world examples from our
home state of Massachusetts to illustrate how these five points of
friction present obstacles to transfer pathways, and offer
recommendations for how they can be best overcome.

As a form of case study, we consider Alice and Bob, two hypothetical
two-year college students interested in data science. Alice attends
Bunker Hill Community College and is pursuing the Data Analytics
associate's degree. Bob attends Holyoke Community College (HCC) and is
pursuing the Mathematics MassTransfer associate's degree. Both have the
goal of obtaining a bachelor's degree in data science from one of the
UMass campuses.

\hypertarget{obstacles-to-transfer-pathways}{%
\subsection{Obstacles to transfer
pathways}\label{obstacles-to-transfer-pathways}}

For purposes of illustration, we focus on the bachelor's program in data
science at the University of Massachusetts at Dartmouth
(\url{https://www.umassd.edu/programs/data-science}). We believe that
this is a helpful example because the program has existed for a number
of years, conforms reasonably well to other curricular guidelines in
data science, and has graduated a number of students since the program
was established. However, most of our analysis and its implications are
not specific to this program, and should be relevant to other programs
that may be proposed in other states.\footnote{While data science lacks
  a national curriculum analogous to those in more established
  disciplines like computer science, mathematics, and statistics, a
  general framework for a bachelor's program in data science is taking
  shape. We have been involved in several efforts to shape such
  curriculum and accreditation guidelines at the national level, most
  notably including De Veaux et al. (2017); National Academies of
  Science, Engineering, and Medicine (2018); Gould et al. (2018). We
  have also been leaders in developing courses (Baumer 2015),
  undergraduate majors, and textbooks (Baumer, Kaplan, and Horton 2021)
  that support instruction in data science. These experiences give us
  some standing to anticipate what the general landscape of bachelor's
  programs in data science will look like over the next ten years. The
  handful of bachelor's programs that currently exist will likely double
  at least once in that time, and one of the major goals of this paper
  is to coordinate efforts across all parties such that the MassTransfer
  system is prepared to adapt to this changing landscape.}

\begin{figure}

{\centering \includegraphics[width=1\linewidth]{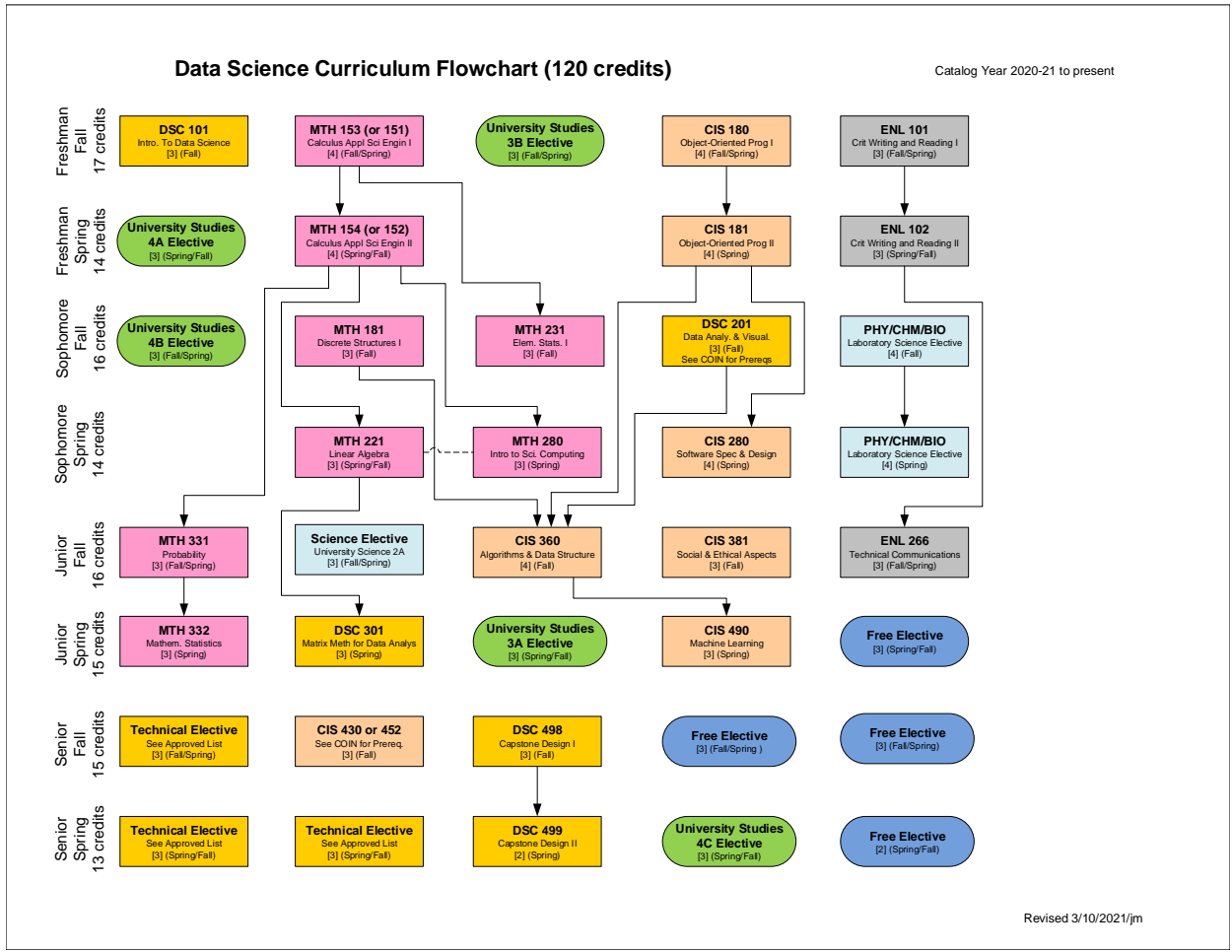} 

}

\caption{Flowchart illustrating progress through the data science bachelor's program at UMass-Dartmouth. For a transfer pathway to work roughly the top half of the flowchart must be completed at a two-year college. Data science courses appear in yellow rectangles.}\label{fig:flowchart}
\end{figure}

Figure \ref{fig:flowchart} illustrates the flow through the eight
semesters of the UMass-Dartmouth data science major. In order for a
transfer pathway to be viable, a student would have to complete the
equivalent of the top half (61 credits) of the flowchart at a two-year
college. Nearly all of the courses coded as mathematics (pink), computer
science (tan), lab sciences (light blue), English (gray), and university
electives (green) have existing mapped equivalents at many two-year
colleges in Massachusetts. However, the two data science courses
(yellow) have no equivalents at the two-year college level.

In Section \ref{sec:proposal}, we use Alice to illustrate why a proposed
pathway from Bunker Hill Community College to UMass-Dartmouth is
necessary, and how it could work.

In this Section, we focus on Bob. Recall that Bob is pursuing the
Mathematics MassTransfer option at HCC. This degree will set him up to
seamlessly transfer to almost all of the public four-year institutions
in Massachusetts, including UMass-Amherst (his first choice) and
UMass-Dartmouth. Bob is from the Holyoke area and wants to stay in
Western Mass; UMass-Amherst is nearby and many of his friends are
already there. It's also the flagship campus and probably comes with the
greatest prestige and employment prospects. The path of least resistance
for Bob would be to transfer to UMass-Amherst and major in mathematics.
However, Bob took the new introduction to data science course at HCC
during his last semester, and now he is interested in data science,
which he believes has more plentiful and lucrative employment options
than mathematics. He knows his background in mathematics will serve him
well, but he needs to learn more about statistics and computer
programming to build out his data science skill set. As of 2022,
UMass-Amherst doesn't offer a full-blown bachelor's degree in data
science, but UMass-Dartmouth does. Bob faces a tough choice:

\begin{itemize}
\tightlist
\item
  If he transfers to UMass-Amherst and pursues a bachelor's in
  mathematics, he probably won't have room in his schedule to flesh out
  his data science skills, and his employment options may suffer.
\item
  If he transfers to UMass-Amherst and then tries to change his major
  (perhaps to informatics with a data science concentration), he might
  not be able to finish his bachelor's degree in two years, costing him
  both time and money.
\item
  If he transfers to UMass-Dartmouth, he's missing several important
  prerequisites, so he wouldn't be able to jump right into the
  junior-year data science curriculum, and risks falling even further
  behind while he struggles to catch up. Plus, he'll be further from the
  support mechanisms he has closer to home.
\item
  If he decides to stay at HCC for another semester (or another year) so
  that he can take computer science classes that are outside of his
  Mathematics MassTransfer degree, he will complicate his financial aid
  situation and likely have to pay out of pocket.
\end{itemize}

Bob is stuck. One reason that Bob is stuck is because he is trying to
change the direction of his course of study midway through his
undergraduate experience. However, it's worth emphasizing that is much
more problematic \emph{because he started at a two-year college}.
Further, if a data science transfer pathway existed, Bob would likely
have a better chance of executing his switch, or at the very least,
might have recognized his interest in data science earlier in his
journey. Thus, we contend that the lack of data science transfer
pathways negatively impacts two-year college students relative to their
four-year college peers.

\hypertarget{sec:dsc101}{%
\subsubsection{Data Science I}\label{sec:dsc101}}

Consistent with the recommendations of De Veaux et al. (2017) and Gould
et al. (2018), students at UMass-Dartmouth take a first course in data
science (Yan and Davis 2019) in the first semester of their first year.
While many other institutions nationally are now teaching introductory
data science, not all students have access at their institutions. It is
not possible to imagine a sensible transfer pathway in which students
are not exposed to the key ideas in data science until their junior
year. Irrespective of pathways to degrees it is critically important
that two-year college students have the opportunity to develop these
skills.

Note that such a course is not simply a grab bag of existing material
from existing courses in statistics and computer science, but rather
focuses on new components of data acumen including the data science
lifecycle, and historically underdeveloped skills like data wrangling
and data visualization that support---but are not subsumed
within---those existing courses. Some courses (e.g.,
\href{http://data8.org/}{Data 8 at Berkeley} and Çetinkaya-Rundel
(2020)) include elements of statistical modeling and inferential
statistics, while others (e.g., SDS 192 at Smith College) do not. In
either case, a first course in statistics is a separate requirement that
exists under many existing transfer pathways in mathematics.

\textbf{In order for data science transfer pathways to work, two-year
colleges must offer a first course in data science.} This is by far the
largest obstacle to bringing these pathways online and the place where
the biggest gain will be achieved in helping institutions to make data
science accessible to their students. This will help to partially
address the recommendation from National Academies of Science,
Engineering, and Medicine (2018) that: ``To prepare their graduates for
this new data-driven era, academic institutions should encourage the
development of a basic understanding of data science in all
undergraduates.''

Many new introductory data science courses will be developed in the
coming years, and it is vital that faculty at the bachelor's and
associate's levels coordinate their efforts to ensure that explicit
course mappings are created that will facilitate transfer.

\begin{quote}
\textbf{Our recommendation} is that institutions develop a flexible,
shared understanding of what constitutes a first course in data science,
and that any new courses developed at any institution are designed with
transfer mappings in mind.
\end{quote}

Ideally, such a first course would:

\begin{enumerate}
\def\labelenumi{\arabic{enumi}.}
\tightlist
\item
  have minimal prerequisites;
\item
  dovetail in useful ways with introductory computer science and
  statistics courses to allow students to take these foundational
  courses in any order;
\item
  transfer to a variety of programs at the bachelor's level, and;
\item
  satisfy a variety of distribution requirements (in Massachusetts, this
  might include the
  \href{https://www.umass.edu/gened/objectives-designations/curricular-designations/basic-mathematics-and-analytic-reasoning}{R2
  analytical reasoning designation at UMass-Amherst}).
\end{enumerate}

At a high level, such a course should prepare students to demonstrate
the ability to:

\begin{itemize}
\tightlist
\item
  use a general-purpose computational environment (e.g., Python or R) to
  analyze data
\item
  scrape, process, clean, and wrangle data from various sources,
  including relational databases
\item
  visualize and interpret relationships between variables in
  multidimensional data
\item
  design accurate, clear, and appropriate data graphics
\item
  communicate the results of an analysis in a correct and comprehensible
  manner
\item
  collaborate within a reproducible workflow
\item
  assess the ethical implications to society of data-based research,
  analyses, and technology in an informed manner.
\end{itemize}

Some courses may choose to cover statistical modeling, while others may
leave that topic in the introductory statistics course. In any case, new
course structures should facilitate an inclusive and engaging learning
environment for students. The Dana Center (2021a) has curated a set of
course design principles that we suggest be incorporated in the course
development process.

Bob is lucky that he took an introductory data science course at HCC.
Most two-year colleges do not yet offer this course.

\hypertarget{sec:dsc201}{%
\subsubsection{Data Science II}\label{sec:dsc201}}

Cultivating a rich facility in data science requires repeated exposure:
a single course is not sufficient for students to develop mastery. To
help students along this path, bachelor's programs in data science
typically include a second course in data science, often taken during
the sophomore year. This course is intended to reinforce and extend
fundamental skills in data wrangling, data visualization, statistical
modeling, and predictive analytics. A richer treatment of data
technologies and database querying in SQL may arise in such a course.
The second course may be taught in a different language (e.g., Python)
than the first course (e.g., R). The focus of the second course will
vary from institution to institution depending on the focus of the first
course (see Section \ref{sec:dsc101}), but we expect the general content
areas to be similar to those listed above. The
\href{https://ds100.org/}{Data 100 course at Berkeley} and the
\href{http://www.cis.umassd.edu/~dkoop/dsc201-2018fa/}{DSC 201} course
at UMass-Dartmouth are examples of second courses in data science.

Second courses in data science obviously depend on a first course, and
often build upon on other core requirements, which may include: a first
course in programming, a first course in statistics, and/or linear
algebra. These prerequisites have an impact on student pathways and may
necessitate delaying completion of this course to the sophomore
year.\footnote{Some two-year college students may need to complete
  additional developmental math courses. Ideally, co-requisite
  approaches (see efforts by the Dana Center (2021a)) could allow them
  to complete these requisites in a timely fashion without delaying
  their progress towards their associate's degree.}

Given the difficulty of launching a first course in data science at
two-year colleges, it may be best, especially in the short-term, to
leave the second course in data science to the universities. While not
optimal, it may be feasible for transfer students to take their second
course in data science during the first semester of their junior year,
and while this will likely disrupt their path relative to non-transfer
students, that disruption can be minimized.

To see how, note that the fifth row in Figure \ref{fig:flowchart},
labelled ``Junior Fall'' lists courses in probability, algorithms \&
data structures, social \& ethical aspects of computing, technical
communications, and a science elective. Some two-year colleges offer
some of these courses. With appropriate planning, transfer students
might be able to take at least one of these courses at their two-year
college in place of a second course in data science, which they would
then take upon transferring. This exchange is possible in part because
data science straddles mathematics and computer science, and data
science students need not complete the entire two-year college
curriculum in \emph{both} mathematics \emph{and} computer science.
Specifically, a computer science class taught in an appropriate language
might help develop their computational foundation and may allow transfer
students to be in a stronger position to excel in their subsequent
courses in data science.

\begin{quote}
\textbf{Our recommendation} is that, for the next few years, second
courses in data science are left to bachelor's programs, and the credits
are replaced with another course with an existing mapping. Planning
should begin on course designs and frameworks for such a course to be
taught at both two- and four-year institutions since this would support
both students planning to transfer as well as associate's-to-workforce
programs.
\end{quote}

Bob hasn't had anything like a second course in data science, so he will
have to take this sophomore-level course as a junior. While this should
end up being a relatively minor detour, it contributes to his feeling of
not quite fitting in with his new classmates (as a two-year college
transfer student). Luckily, Bob used one of his general electives to
take a programming course that prepares him for CIS 360.

\hypertarget{sec:workflow}{%
\subsubsection{A course in scientific computing, data science workflow,
and reproducible computing}\label{sec:workflow}}

A generic bachelor's program in data science will include explicit
instruction in how to advance science by computing with data in a
reproducible, collaborative workflow. In some programs, this instruction
will be woven into modules that permeate a series of courses. In others,
there will be a standalone course that focuses on these issues. It is
important that the technologies to support workflow and reproducible
analysis as a component of data acumen (National Academies of Science,
Engineering, and Medicine 2018) should not be assumed to be known by
students or left for them to learn outside of a course, lest existing
disparities in background are exacerbated.

Topics in this area include version control systems (e.g.,
\texttt{git}), collaboration and project management tools (e.g., GitHub,
Trello), software development paradigms (e.g., Agile/Scrum), document
authoring software (e.g., Pandoc, variants of \texttt{markdown}
(Jupyter, R Markdown, Quarto), \LaTeX), command line scripting (e.g.,
UNIX), cloud computing, as well as further exposure to R, Python, and/or
SQL.

While there are existing models of such courses at two-year colleges,
they are less likely to have existing transfer mappings. Given the
variety of topics in these courses and the difficulty of coordinating
the content across institutions, these credits will probably have to be
mapped on a one-to-one basis. One promising avenue is a course in R or
Python that is outside of the main computer science sequence (which is
often taught in Java or C++). An example of such a course is CSE 160 at
Springfield Technical Community College.

\begin{quote}
\textbf{Our recommendation} is that individual programs map credits
where reasonably equivalent options exist, and replace them with general
education or liberal arts credits where they don't.
\end{quote}

Bob was under-prepared for the mechanics of data science at the
junior-level. While he did use R Markdown in his data science course, he
had never used GitHub, SQL, or the command line. The lack of such
experience made it more difficult for him to secure an internship during
his work on his associate's degree. Other frustrations after
transferring occurred when he felt like he understood the material in
his classes but struggled to participate in group projects because he
wasn't as facile with the workflow tools. Two- and four-year
institutions should work to provide flexible options for students to be
introduced to and deepen their understandings of reproducibility and
workflow throughout their courses and programs (Horton et al. 2022).

\hypertarget{sec:lab}{%
\subsubsection{Lab sciences}\label{sec:lab}}

Many of the existing Massachusetts transfer options in computer science
(and other STEM disciplines) require two semesters of lab sciences
(e.g., physics, biology, or chemistry) as a component of their general
education requirements. Requiring a student pursuing a bachelor's degree
in data science to take two semesters of physics, biology, or chemistry
provides an opportunity for them to learn important aspects of the
scientific process as well as the collection and analysis of data. At
present, many of these courses may be less germane for data science
students, but there is considerable potential for them to reinforce and
build basic data sciences skills for all students while building domain
knowledge.

As an alternative to explore, we can imagine that a future data science
infused lab course could be developed as a way to provide more exposure
to key data science topics while meeting the learning outcomes for a lab
course. This is perhaps less of a ``friction-point'' than an opportunity
to improve data science options as well as to infuse computation and
data into undergraduate STEM education (National Academies of Science,
Engineering, and Medicine 2022).

\begin{quote}
\textbf{Our recommendation} is that students use existing pathways for
lab sciences, choosing courses when possible that incorporate aspects of
scientific data (e.g., Greenfield Community College's
\href{https://my.gcc.mass.edu/PROD/bwckctlg.p_disp_course_detail?cat_term_in=202201\&subj_code_in=BIO\&crse_numb_in=120}{BIO
120 Introduction to Environmental Science})\footnote{Shodor
  (\url{https://http://shodor.org}) has worked to incorporate data
  science into various STEM majors, including biology, chemistry, and
  physics.}.
\end{quote}

Bob completed this sequence as part of his transfer degree.

\hypertarget{communication-ethics-and-application-domains}{%
\subsubsection{Communication, ethics, and application
domains}\label{communication-ethics-and-application-domains}}

Bachelor's programs in data science include training in communication
(how do we transfer knowledge gained from data analysis from data
scientist to a broader audience? (Parke 2008)) and ethics (what
responsibilities do data scientists have to their users, customers, and
society as a whole? (Baumer et al. 2022)). In addition, a domain of
application is valuable (how does data science enhance our understanding
of another subject?). These vital aspects of a data science curriculum
cannot wait entirely until the junior year, and thus, two-year college
students must find ways to build skills in these areas before they
transfer.

Most two-year colleges offer courses in communication. If any of those
courses focus on \emph{communicating with data}, they should be taken.
Courses that focus on more general writing skills are still valuable,
and are already part of the general education requirements for any
associate's degree. Where courses in ethics, or preferably, data ethics
are available, they should be taken at the two-year college level, as
this will help to infuse ethics early in a student's education.

For those students whose application domain will intersect with the lab
sciences mentioned in Section \ref{sec:lab}, that requirement might
provide a helpful synergy. We imagine that this might be particularly
beneficial for students interested in public health, biostatistics, or
bioinformatics.

One challenge here will be ensuring that whatever these courses are,
they count towards the associate's degree program.

\begin{quote}
\textbf{Our recommendation} is that institutions think carefully and
holistically about how requirements for communication, ethics, and
domain application can be used to accrue credits at two-year colleges
and foster successful transfers.
\end{quote}

Although Bob was exposed to ideas in data science ethics during his
course at HCC, there wasn't time for much depth. Thus, he feels as
though he is learning about these issues in depth for the first time in
his junior year. He wishes he had thought to take CSI 215 (Ethical/Legal
Aspects of Information Systems) at HCC, since it already transfers as
CIS 381 at UMD, which he now needs to take anyway.

\hypertarget{where-to-situate-programs}{%
\subsection{Where to situate
programs?}\label{where-to-situate-programs}}

Unfortunately, the interdisciplinary nature of data science is in
conflict with the siloing of programs within departments. The 2018 NASEM
data science for undergraduates report found that many bachelor's degree
programs in data science are housed in a college or school of business,
a mathematics or statistics department, or a computer science department
(see pages 3--5 of National Academies of Science, Engineering, and
Medicine (2018)). A few undergraduate data science majors were described
as hybrids of these three models, with joint administration/programmatic
coordination. We believe that such hybrid models are better suited to
ensure that students develop a deep foundation in all aspects of data
acumen.

When considering where to situate associate's degree programs within
departments at two-year colleges, the compressed timeline given the
two-year nature of the degree only compounds the problem. As a result,
until there are associate's degree programs in data science, even
explicit transfer pathways (such as the ones we are trying to create)
may force students to choose between two potentially undesirable
options: obtaining an associate's degree in liberal arts studies that
may not be as marketable as a degree in a more technical field, or
supplementing a degree in mathematics or computer science with several
additional courses. Our hope is to provide guidance about flexible
pathways that could soften these rough edges that exist at present.

\hypertarget{alignment-of-degrees-with-pathways}{%
\subsection{Alignment of degrees with
pathways}\label{alignment-of-degrees-with-pathways}}

Even without the kind of explicit pathways we are advocating for,
transfer to a bachelor's in data science may still be possible. However,
a student would have to forge their own pathway, which might mean taking
courses at a two-year college that were outside of the requirements of
their associate's degree program, taking catch-up courses at a four-year
college once they arrive, and/or obtaining explicit transfer credit for
courses that are not already mapped. \textbf{All of these obstacles add
unnecessary friction, cost, and time that students and society cannot
afford.}

\hypertarget{sec:proposal}{%
\section{Case study: A new transfer pathway in data
science}\label{sec:proposal}}

In this Section, we articulate a realistic vision for a new transfer
pathway from Bunker Hill Community College to UMass-Dartmouth (see
Figure \ref{fig:flowchart}). We believe that such a pathway, if
approved, would be the first of its kind\footnote{In the fall of 2021,
  Bunker Hill Community College signed an articulation agreement with
  Northeastern University from their associate of science in data
  analytics degree to Northeastern's bachelor of science in analytics
  degree program. This agreement serves as a proof-of-concept that
  transfer pathways to \emph{public} universities in Massachusetts can
  be created.} in Massachusetts, and one of only a handful in the
nation. We hope that this proposal will serve as a model for similar
potential pathways between other two-year colleges and universities.

\hypertarget{bunker-hill-to-umass-dartmouth}{%
\subsection{Bunker Hill to
UMass-Dartmouth}\label{bunker-hill-to-umass-dartmouth}}

Bunker Hill Community College offers a
\href{https://catalog.bhcc.edu/preview_program.php?poid=62}{data
analytics option} within the associate of science program. Students in
this program take multiple courses in computer science, receive
foundational training in statistics, linear algebra, and college
writing, and are exposed to R, Python, and SQL. This is a terminal
degree which is not designed for transfer. So while most of the pieces
for a data science transfer pathway from Bunker Hill Community College
(BHCC) to UMass-Dartmouth are already in place, the degree programs do
not align.

Although students who attend BHCC may be more interested in staying in
Boston and transferring to Northeastern via an existing articulation
agreement, a public option at UMD would be the first potential
MassTransfer pathway in data science. Tables \ref{tab:bhcc-umd-inner},
\ref{tab:bhcc-umd-left}, \ref{tab:bhcc-umd-right} illustrate the current
situation. Most of the course mappings in these figures are already
approved by the MassTransfer system. In what follows, we provide detail
about the exceptions, and discuss possibilities to reduce the number of
``wasted'' (or ``stranded'') credits.

\begin{table}

\caption{\label{tab:unnamed-chunk-4}Courses that should easily transfer between Bunker Hill Community College's data analytics associate's degree program and UMass-Dartmouth's data science bachelor's degree program. These courses count towards both degrees and most mappings are already approved by the MassTransfer system. Note that 5 credits are `lost' in the process. The `Mapped' column indicates whether the course mapping is already approved in the MassTransfer system. \label{tab:bhcc-umd-inner}}
\centering
\begin{tabu} to \linewidth {>{\raggedright\arraybackslash}p{2in}>{\raggedright}X>{\raggedleft\arraybackslash}p{3em}>{\raggedright}X>{\raggedleft\arraybackslash}p{3em}>{\raggedright\arraybackslash}p{3em}}
\toprule
Topic & BHCC & BHCC Credit & UMD & UMD Credit & Mapped\\
\midrule
\cellcolor{gray!6}{Calculus I} & \cellcolor{gray!6}{MAT 281} & \cellcolor{gray!6}{4} & \cellcolor{gray!6}{MTH 151} & \cellcolor{gray!6}{4} & \cellcolor{gray!6}{Y}\\
Calculus II & MAT 282 & 4 & MTH 152 & 4 & Y\\
\cellcolor{gray!6}{Linear Algebra} & \cellcolor{gray!6}{MAT 291} & \cellcolor{gray!6}{4} & \cellcolor{gray!6}{MTH 221} & \cellcolor{gray!6}{3} & \cellcolor{gray!6}{Y}\\
Statistics \& Data Science & CIT 130 + CIT 137 + CIT 187 & 10 & DSC 101 + DSC 201 + MTH 231 & 9 & N\\
\cellcolor{gray!6}{Scientific Computation} & \cellcolor{gray!6}{CSC 125 + CSC 225} & \cellcolor{gray!6}{6} & \cellcolor{gray!6}{MTH 280} & \cellcolor{gray!6}{3} & \cellcolor{gray!6}{N}\\
\addlinespace
College Writing I & ENG 111 & 3 & ENL 101 & 3 & Y\\
\cellcolor{gray!6}{College Writing II} & \cellcolor{gray!6}{ENG 112} & \cellcolor{gray!6}{3} & \cellcolor{gray!6}{ENL 102} & \cellcolor{gray!6}{3} & \cellcolor{gray!6}{Y}\\
Scientific Reasoning & multiple & 4 & NA & 4 & Y\\
\cellcolor{gray!6}{Creative Work} & \cellcolor{gray!6}{multiple} & \cellcolor{gray!6}{3} & \cellcolor{gray!6}{NA} & \cellcolor{gray!6}{3} & \cellcolor{gray!6}{Y}\\
Community \& Cultural Contexts & multiple & 3 & NA & 3 & Y\\
\addlinespace
\cellcolor{gray!6}{Excel} & \cellcolor{gray!6}{CIT 234} & \cellcolor{gray!6}{3} & \cellcolor{gray!6}{NA} & \cellcolor{gray!6}{3} & \cellcolor{gray!6}{Y}\\
\midrule
Total & - & 47 & - & 42 & -\\
\bottomrule
\end{tabu}
\end{table}

\begin{table}

\caption{\label{tab:unnamed-chunk-5}Courses that are required for the data analytics associate's degree at Bunker Hill Community College, but do not count towards the bachelor's degree at UMass-Dartmouth. These credits are `wasted' in the sense that a two-year college student needs them to graduate, but not to transfer. \label{tab:bhcc-umd-left}}
\centering
\begin{tabu} to \linewidth {>{\raggedright\arraybackslash}p{2in}>{\raggedright}X>{\raggedleft\arraybackslash}p{3em}>{\raggedright}X>{\raggedleft\arraybackslash}p{3em}>{\raggedright\arraybackslash}p{3em}}
\toprule
Topic & BHCC & BHCC Credit & UMD & UMD Credit & Mapped\\
\midrule
\cellcolor{gray!6}{Intro Statistics} & \cellcolor{gray!6}{MAT 181} & \cellcolor{gray!6}{3} & \cellcolor{gray!6}{MTH 147} & \cellcolor{gray!6}{3} & \cellcolor{gray!6}{Y}\\
Problem Solving & CIT 113 & 3 & NA & NA & N\\
\cellcolor{gray!6}{Precalculus} & \cellcolor{gray!6}{MAT 197} & \cellcolor{gray!6}{4} & \cellcolor{gray!6}{MTH 150} & \cellcolor{gray!6}{3} & \cellcolor{gray!6}{Y}\\
SQL Programming & CSC 236 & 3 & NA & NA & N\\
\cellcolor{gray!6}{Database Programming} & \cellcolor{gray!6}{CSC 240} & \cellcolor{gray!6}{3} & \cellcolor{gray!6}{NA} & \cellcolor{gray!6}{NA} & \cellcolor{gray!6}{N}\\
\addlinespace
Operating Systems & CIT 268 & 3 & NA & NA & N\\
\midrule
\cellcolor{gray!6}{Total} & \cellcolor{gray!6}{-} & \cellcolor{gray!6}{19} & \cellcolor{gray!6}{-} & \cellcolor{gray!6}{6} & \cellcolor{gray!6}{-}\\
\bottomrule
\end{tabu}
\end{table}

\begin{table}

\caption{\label{tab:unnamed-chunk-6}Courses that are required for the data science bachelor's program at UMass-Dartmouth but not for the data analytics associate's degree at Bunker Hill Community College. Two-year college students might have to pay extra in order to satisfy these additional requirements in order to transfer. \label{tab:bhcc-umd-right}}
\centering
\begin{tabu} to \linewidth {>{\raggedright\arraybackslash}p{2in}>{\raggedright}X>{\raggedleft\arraybackslash}p{3em}>{\raggedright}X>{\raggedleft\arraybackslash}p{3em}>{\raggedright\arraybackslash}p{3em}}
\toprule
Topic & BHCC & BHCC Credit & UMD & UMD Credit & Mapped\\
\midrule
\cellcolor{gray!6}{Physics I} & \cellcolor{gray!6}{PHY 251} & \cellcolor{gray!6}{4} & \cellcolor{gray!6}{PHY 113} & \cellcolor{gray!6}{4} & \cellcolor{gray!6}{Y}\\
Physics II & PHY 252 & 4 & PHY 114 & 4 & Y\\
\cellcolor{gray!6}{Programming I} & \cellcolor{gray!6}{CSC 239} & \cellcolor{gray!6}{4} & \cellcolor{gray!6}{CIS 180} & \cellcolor{gray!6}{4} & \cellcolor{gray!6}{Y}\\
Programming II & CSC 285 & 4 & CIS 181 & 4 & Y\\
\cellcolor{gray!6}{Discrete Math} & \cellcolor{gray!6}{NA} & \cellcolor{gray!6}{NA} & \cellcolor{gray!6}{MTH 181} & \cellcolor{gray!6}{3} & \cellcolor{gray!6}{N}\\
\addlinespace
Software Design & NA & NA & CSC 280 & 4 & N\\
\midrule
\cellcolor{gray!6}{Total} & \cellcolor{gray!6}{-} & \cellcolor{gray!6}{16} & \cellcolor{gray!6}{-} & \cellcolor{gray!6}{23} & \cellcolor{gray!6}{-}\\
\bottomrule
\end{tabu}
\end{table}

Alice is interested in pursuing this pathway, but because it doesn't
exist, she can't automatically enroll in the data science program at
UMD. By completing the data analytics degree program at BHCC, she will
complete all of the courses listed in Tables \ref{tab:bhcc-umd-inner}
and \ref{tab:bhcc-umd-left}. This gives her 63 credits, which is more
than the 60 credits typically required for a transfer pathway.
Unfortunately, only 36 of those 63 credits are transferable to UMD, and
because Alice is missing all of the courses in Table
\ref{tab:bhcc-umd-right}, she's not actually prepared for the
junior-level data science curriculum at UMass-Dartmouth. She could pay
out of pocket for the 16 transferable credits in Table
\ref{tab:bhcc-umd-right}, but that would cost her time and money, and
still leave her a few requirements short. (Financial aid may also not be
available to support her completion of these ``stranded'' credits.)
Thus, Alice's academic progress may also be impeded.

\hypertarget{a-proposed-pathway}{%
\subsection{A proposed pathway}\label{a-proposed-pathway}}

Tables \ref{tab:bhcc-umd-inner}, \ref{tab:bhcc-umd-left},
\ref{tab:bhcc-umd-right} lay out the building blocks from which an
articulation agreement in data science between BHCC and UMD. We argue
that this articulation agreement would provide a proof-of-concept and
blueprint for a more generalized \emph{data science MassTransfer
associate's degree}. The program would consist of all the courses in
Tables \ref{tab:bhcc-umd-inner} and \ref{tab:bhcc-umd-right}, with
perhaps a few tweaks.

We note first that most of the courses in Tables
\ref{tab:bhcc-umd-inner} and \ref{tab:bhcc-umd-right} are already
mapped. The exceptions in Table \ref{tab:bhcc-umd-inner} are the
statistics, data science, and scientific computing blocks. Creating
these course mappings is an active conversation among the authors and
the relevant parties. We're optimistic that these mappings will remove
the first, second, and third points of friction, although the student
stands to lose five credits in the process.

Among the courses in Table \ref{tab:bhcc-umd-right}, four are already
mapped and would be part of the new transfer pathway. The physics
courses remove the fourth point of friction. Two courses (discrete math
and software design) have no obvious equivalent at BHCC. Discrete math
is taught at many two-year colleges, and so the lack of a discrete math
option at BHCC is more an idiosyncracy of BHCC than a systemic problem.
We hope that a suitable alternative course can be found, or that UMD
will accept an elective in its place. As for the software design course,
BHCC offers a number of alternative courses that, while not covering the
same material, seem like they would prepare future data scientists
equally well, albeit in different ways. For example, the two course
sequence in SQL Programming and Database Programming shown in Table
\ref{tab:bhcc-umd-left} seems like it covers highly relevant material.
Perhaps UMD would accept those courses as alternatives.

In that event, suppose that we substitute the six credit sequence in SQL
programming for the discrete math and software design requirement at
UMD. Then adding these credits to those in Tables
\ref{tab:bhcc-umd-inner} and \ref{tab:bhcc-umd-right} results in a 69
credit curriculum of existing courses at BHCC, that would transfer as 64
credits that cover nearly all of the first two years of the data science
curriculum at UMD.

Using only courses that already exist, we see this as the best candidate
to be the first public data science transfer pathway in Massachusetts.

\hypertarget{sec:conclusion}{%
\section{Conclusion}\label{sec:conclusion}}

At the December 2018 meeting of the National Academies Postsecondary
Data Science Education Roundtable (National Academies of Sciences,
Engineering, and Medicine 2020), D.J. Patil, former Chief Data Scientist
in the White House Office of Science and Technology Policy, described
how his experience from a two-year college bestowed upon him ``three
gifts'': ``a love of mathematics, an understanding of how to write in
various genres, and confidence to succeed at the postsecondary level
(page 158)''. He expressed that his experience at two-year college
provided a crucial ``on-ramp'' to his future success in data science.

Like Patil, we see two-year colleges as key players in developing the
next generation of data science students. Our experience with the
DSC-WAV project and other interactions have shown that our two-year
college system has countless committed and engaged educators and
administrators working to build better futures for their students,
amidst time and resource constraints. Patil testifies to the
habits-of-mind and general skills that two-year colleges are already
cultivating---our goal is to align curricula so as to reduce
administrative and bureaucratic obstacles.

We have focused on associate's degrees that prepare for a bachelor's
degree. Other pathways, such associate's-to-workforce are also
important, and need improved flexibility and transparency.

\hypertarget{additional-challenges}{%
\subsection{Additional challenges}\label{additional-challenges}}

There is considerable work needed to foster sustainable courses,
structures, and programs. We acknowledge that this will require focus
and attention for many years. Efforts such as the NSF-funded EDC Oceans
of Data ``Mentoring New Data Pathways'' project
(\url{http://oceansofdata.org/projects/mentoring-new-data-pathways-community-colleges})
have engaged Bunker Hill Community College in an effort to support new
data programs.

Resource disparities at many two-year colleges and insufficient
partnerships between two- and four-year institutions could hamper these
efforts. As but one example, due to resources and other circumstances,
two-year colleges could feel at a disadvantage and perhaps be reluctant
to offer courses that are not included in guaranteed transfer systems
such as MassTransfer, or courses not belonging to an already structured
pathway.

Faculty development is another critical issue. At a time when data
science positions are challenging for employers to fill, where will the
next generation of instructors come from? This is another area where
partnerships between two- and four-year institutions as well as industry
will be critical (National Academies of Science, Engineering, and
Medicine 2018). An example of how this might work for data science can
be seen in the strategies outlined in A. Enriquez et al. (2018) for
engineering transfer programs.

The changing preK-12 landscape raises important questions. As states are
reviewing and revising their mathematics, science, and computing
standards, statistics and data science are being elevated and made more
explicit. We believe that this will impact the knowledge, skills, and
abilities students bring to their post-secondary education. These
changes may impact the future of pathways, potentially in positive ways.

A reviewer noted that one of the biggest challenges to students
transferring to STEM programs is that they get caught in a ``mathematics
maze.'' A failed course leads to remedial courses which lead to more
courses, which impedes progress towards completion of their program. We
agree that this is an important unsolved issue (see Cafarella (2021) for
an in-depth treatment). In other disciplines (e.g., nursing), efforts
such as the Dana Center's Mathematics Education for Nurses collaboration
(Dana Center 2021b) works to improve student success while helping
nursing students gain the ``mathematical knowledge, skills, and
attitudes'' to be successful in their career. Similar efforts would
benefit future data science programs.

There are many other issues that we could address at this juncture,
including aspects of associate's to workforce programs, challenges and
opportunities of dual enrollment, and the pressing need for improved
computational infrastructure. But we intentionally limit our primary
focus to fostering pathways, which needs to begin by identifying
barriers and resources to the widespread teaching of accessible and
pedagogically sound introductory data science courses.

\hypertarget{bright-spots}{%
\subsection{Bright spots}\label{bright-spots}}

In addition to the proof-of-concept at UNH-Manchester, there are some
useful models that we can consider.

\href{https://data.berkeley.edu/californiaalliance}{Considerable
efforts} towards pathways are underway in California. While the scale of
the California system---which includes both the UC and Cal State
constellations---provides obvious challenges, there have been
encouraging developments. Models for addressing similar challenges in
related disciplines (e.g., engineering) exist (A Enriquez et al. 2018).

In Ohio, 36 public institutions of higher education, 27 two-year
colleges, and 9 four-year colleges approved a set of learning outcomes
for a general education data science course developed by faculty from
two- and four-year institutions (Ricardo Moena, personal communication).
We see this as a necessary but not sufficient step.

In Connecticut, recent efforts have led to the establishment of
\href{https://www.nwcc.edu/completion/as-data-science/}{associate's
programs in data science} with pathways to workforce or transfer
(Northwestern CT Community College 2020).

\hypertarget{a-call-to-broaden-participation}{%
\subsection{A call to broaden
participation}\label{a-call-to-broaden-participation}}

We close with some reflections on the critical role that two-year
colleges provide in terms of \textbf{affordable} options that are
accessible to a \textbf{diverse} population.

The Broadening Data Science Education (Rawlings-Goss et al. 2018) report
notes that:

\begin{quote}
Many individuals in today's data science workforce are coming from
doctoral or master's degree programs, which have seen a dramatic
increase in recent years. While these advanced degrees are valuable, it
is not economically feasible for all data scientists to complete four
years of an undergraduate degree, then a one- or two-year master's
program before they can undertake useful work. Ensuring the future
growth of the workforce requires an expansion to four-year and two-year
degrees (page 45).
\end{quote}

At the June 2019 NASEM Roundtable meeting, Uri Treisman of the
University of Texas-Austin and the Dana Center described data science
programs as ``powerful resources for students seeking upward mobility
(page 165).'' (National Academies of Sciences, Engineering, and Medicine
2020)

Moreover, the Broadening Data Science Education (Rawlings-Goss et al.
2018) report suggests that: ``the potential impact of the Data Divide is
no less dire for our institutions of higher education'' (page 7). Such
concerns lead to the finding that: ``Data science would particularly
benefit from broad participation by underrepresented minorities because
of the many applications to problems of interest to diverse
populations.'' (National Academies of Science, Engineering, and Medicine
2018) The \href{https://data.berkeley.edu/californiaalliance}{California
Alliance for Data Science Education} notes that ``increasing access to
data science as a career option for all students is key to making data
science a more diverse and inclusive field.'' The Broadening Data
Science Education (Rawlings-Goss et al. 2018) report states this even
more directly:

\begin{quote}
If we do not make diversity and inclusion a priority now, we will not
have it in the future. We do not want to repeat the mistakes of the
past, so we must reverse the trend for the growing divide to make and
keep data science broad. Diversity will bring a lot of ideas and voices
to the table, which may lead to significantly fewer models producing
biased results when trained using algorithms on biased data sets. (page
30).
\end{quote}

We agree that two-year colleges are the only affordable game in town and
serve a key role in data science education now and in the future.

\hypertarget{acknowledgements}{%
\section{Acknowledgements}\label{acknowledgements}}

We acknowledge the many efforts of DSC-WAV Project Coordinator Andrea
Dustin, our many collaborators and students on the project, as well as
financial support from NSF grants HDR DSC-1923388 and HDR DSC-1924017.
We appreciate the input and efforts of the co-PIs from our local
two-year colleges: Ileana Vasu (Holyoke), Ebenezer Afarikumah
(Greenfield), and Brian Candido (Springfield Technical). We thank Brant
Cheikes, Matthew Rattigan, Tom Bernadin, Michelle Trim, Scott Field, and
Iren Valova for sharing their thoughts and suggestions. Sarah Dunton,
Jenn Halbleib, Michael Harris, Tyler Kloefkorn, Kate Kozak, Donna
LaLonde, Sears Merritt, Ricardo Moena, Roxy Peck, Josh Recio, Rachel
Saidi, and Rebecca Wong provided many helpful comments and suggestions
on an earlier draft of the manuscript.

\hypertarget{references}{%
\section{References}\label{references}}

\hypertarget{refs}{}
\begin{cslreferences}
\leavevmode\hypertarget{ref-new_two}{}%
Amstat News. 2022. ``New Two-Year College Data Science, Analytics
Programs on the Rise.'' \emph{Amstat News}, August.
\url{https://magazine.amstat.org/blog/2022/08/01/new-two-year-programs}.

\leavevmode\hypertarget{ref-baumer2015data}{}%
Baumer, Benjamin S. 2015. ``A Data Science Course for Undergraduates:
Thinking with Data.'' \emph{The American Statistician} 69 (4): 334--42.
\url{https://doi.org/10.1080/00031305.2015.1081105}.

\leavevmode\hypertarget{ref-baumerethics}{}%
Baumer, Benjamin S., Randi L. Garcia, Albert Y. Kim, Katherine M.
Kinnaird, and Miles Q. Ott. 2022. ``Integrating Data Science Ethics into
an Undergraduate Major: A Case Study.'' \emph{Journal of Statistics and
Data Science Education} 30 (1).
\url{https://doi.org/10.1080/26939169.2022.2038041}.

\leavevmode\hypertarget{ref-baumer2020mdsr}{}%
Baumer, Benjamin S., Daniel T. Kaplan, and Nicholas J. Horton. 2021.
\emph{Modern Data Science with R}. 2nd ed. Chapman; Hall/CRC Press: Boca
Raton. \url{https://mdsr-book.github.io/mdsr2e}.

\leavevmode\hypertarget{ref-blumenstyk2021}{}%
Blumenstyk, Goldie. 2021. ``The Edge: The `Dirty Secret' That Obstructs
Transfer.'' The Chronicle of Higher Education.
\url{https://www.chronicle.com/newsletter/the-edge/2021-11-10}.

\leavevmode\hypertarget{ref-cafarella2021breaking}{}%
Cafarella, Brian. 2021. \emph{Breaking Barriers: Student Success in
Community College Mathematics}. CRC Press: Boca Raton.
\url{https://www.routledge.com/Breaking-Barriers-Student-Success-in-Community-College-Mathematics/Cafarella/p/book/9781032007977}.

\leavevmode\hypertarget{ref-cleveland2001data}{}%
Cleveland, William S. 2001. ``Data Science: An Action Plan for Expanding
the Technical Areas of the Field of Statistics.'' \emph{International
Statistical Review} 69 (1): 21--26.
\url{https://doi.org/10.1111/j.1751-5823.2001.tb00477.x}.

\leavevmode\hypertarget{ref-cetinkaya2021box}{}%
Çetinkaya-Rundel, Mine. 2020. ``Data Science in a Box.''
https://datasciencebox.org/. \url{https://datasciencebox.org/}.

\leavevmode\hypertarget{ref-DanaCenter}{}%
Dana Center. 2021a. ``Data Science Course Framework.'' 2021.
\url{https://www.utdanacenter.org/sites/default/files/2021-05/data/_science/_course/_framework/_2021/_final.pdf}.

\leavevmode\hypertarget{ref-dana_nurses}{}%
---------. 2021b. ``Mathematics Education for Nurses.'' 2021.
\url{https://www.utdanacenter.org/our-work/higher-education/collaborations/math-for-nurses}.

\leavevmode\hypertarget{ref-danyluk2019computing}{}%
Danyluk, A, P Leidig, S Buck, L Cassel, A McGettrick, W Qian, C Servin,
and H Wang. 2021. ``Computing Competencies for Undergraduate Data
Science Curricula.'' Association for Computing Machinery; Association
for Computing Machinery.
\url{https://dstf.acm.org/DSTF/_Final/_Report.pdf}.

\leavevmode\hypertarget{ref-pcmi2016guidelines}{}%
De Veaux, Richard D., Mahesh Agarwal, Maia Averett, Benjamin S. Baumer,
Andrew Bray, Thomas C. Bressoud, Lance Bryant, et al. 2017. ``Curriculum
Guidelines for Undergraduate Programs in Data Science.'' \emph{Annual
Review of Statistics and Its Application} 4 (1): 1--16.
\url{https://doi.org/10.1146/annurev-statistics-060116-053930}.

\leavevmode\hypertarget{ref-donoho201750}{}%
Donoho, David. 2017. ``50 Years of Data Science.'' \emph{Journal of
Computational and Graphical Statistics} 26 (4): 745--66.
\url{https://doi.org/10.1080/10618600.2017.1384734}.

\leavevmode\hypertarget{ref-statista}{}%
Duffin, Erin. 2021. ``Community Colleges in the United States:
Statistics \& Facts.'' Statista.
\url{https://www.statista.com/topics/3468/community-colleges-in-the-united-states/\#dossierContents__outerWrapper}.

\leavevmode\hypertarget{ref-osti_10063235}{}%
Enriquez, A., N. Langhoff, E. Dunmire, T Rebold, and W. Pong. 2018.
``Strategies for Developing, Expanding, and Strengthening Community
College Engineering Transfer Programs.'' \emph{American Society for
Engineering Education} 2018 (June).
\url{https://par.nsf.gov/biblio/10063235}.

\leavevmode\hypertarget{ref-enriquez2018strategies}{}%
Enriquez, A, Nicholas Langhoff, E Dunmire, Thomas Rebold, and Wenshen
Pong. 2018. ``Strategies for Developing, Expanding, and Strengthening
Community College Engineering Transfer Programs.'' In \emph{American
Society for Engineering Education}, 2018:16.
\url{https://doi.org/10.18260/1-2--30995}.

\leavevmode\hypertarget{ref-gould2018two}{}%
Gould, Rob, R Peck, J Hanson, N J Horton, Brian Kotz, K Kubo, J
Malyn-Smith, et al. 2018. ``The Two-Year College Data Science Summit.''
American Statistical Association.
\url{https://www.amstat.org/asa/files/pdfs/2018TYCDS-Final-Report.pdf}.

\leavevmode\hypertarget{ref-jsdse2022}{}%
Horton, Nicholas J., Rohan Alexander, Aneta Piekut, and Colin Rundel.
2022. ``The Growing Importance of Reproducibility and Responsible
Workflow in the Data Science and Statistics Curriculum.'' \emph{Journal
of Statistics and Data Science Education} 30 (3).
\url{https://doi.org/10.1080/26939169.2022.2141001}.

\leavevmode\hypertarget{ref-horton2021dscwav}{}%
Horton, Nicholas J., Benjamin S. Baumer, Andrew Zieffler, and Valerie
Barr. 2021. ``The Data Science Corps Wrangle-Analyze-Visualize Program:
Building Data Acumen for Undergraduate Students.'' \emph{Harvard Data
Science Review} 3 (1): 1--8.
\url{https://doi.org/10.1162/99608f92.8233428d}.

\leavevmode\hypertarget{ref-korn2021}{}%
Korn, Melissa, and Andrea Fuller. 2021. ``\,`Financially Hobbled for
Life': The Elite Master's Degrees That Don't Pay Off.'' The Wall Street
Journal.
\url{https://www.wsj.com/articles/financially-hobbled-for-life-the-elite-masters-degrees-that-dont-pay-off-11625752773}.

\leavevmode\hypertarget{ref-legacy2021dscwav}{}%
Legacy, Chelsey, Andrew Zieffler, Benjamin S. Baumer, Valerie Barr, and
Nicholas J. Horton. 2022. ``Facilitating Team-Based Data Science:
Lessons Learned from the DSC-WAV Project.'' \emph{Foundations of Data
Science}. \url{https://doi.org/10.3934/fods.2022003}.

\leavevmode\hypertarget{ref-murnane2022massinc}{}%
Murnane, Richard J., John B. Willett, John P. Papay, Ann Mantil, Preeya
P. Mbekeani, and Aubrey McDonough. 2022. ``Building Stronger Community
College Transfer Pathways: Evidence from Massachusetts.'' Massachusetts
Institute for a New Commonwealth.
\url{https://massinc.org/research/building-stronger-community-college-transfer-pathways/}.

\leavevmode\hypertarget{ref-nasem2018}{}%
National Academies of Science, Engineering, and Medicine. 2018.
\emph{Data Science for Undergraduates: Opportunities and Options}.
National Academies Press: Washington, DC.
\url{https://nas.edu/envisioningds}.

\leavevmode\hypertarget{ref-nasemundergrad}{}%
---------. 2022. \emph{Imagining the Future of Undergraduate STEM
Education}. National Academies Press: Washington, DC.
\url{https://nap.nationalacademies.org/read/26314}.

\leavevmode\hypertarget{ref-dsert2020}{}%
National Academies of Sciences, Engineering, and Medicine. 2020.
``Roundtable on Data Science Postsecondary Education.'' American
Statistical Association. \url{https://www.nap.edu/25804}.

\leavevmode\hypertarget{ref-NAP21739}{}%
National Academy of Engineering and National Academies of Sciences,
Engineering, and Medicine. 2016. \emph{Barriers and Opportunities for
2-Year and 4-Year Stem Degrees: Systemic Change to Support Students'
Diverse Pathways}. Edited by Shirley Malcom and Michael Feder.
Washington, DC: The National Academies Press.
\url{https://doi.org/10.17226/21739}.

\leavevmode\hypertarget{ref-NCCC}{}%
Northwestern CT Community College. 2020. ``Northwestern CT Community
College Launches Data Science Degree Program.'' New Haven Register.
\url{https://www.nhregister.com/news/article/Northwestern-CT-Community-College-launches-data-15074344.php}.

\leavevmode\hypertarget{ref-parkecommunication}{}%
Parke, Carol S. 2008. ``Reasoning and Communicating in the Language of
Statistics.'' \emph{Journal of Statistics Education} 16 (1).
\url{https://doi.org/10.1080/10691898.2008.11889555}.

\leavevmode\hypertarget{ref-broadening2018}{}%
Rawlings-Goss, R, L Cassel, M Cragin, C Cramer, A Dingle, S
Friday-Stroud, A Herron, et al. 2018. ``Keeping Data Science Broad:
Negotiating the Digital and Data Divide Among Higher Education
Institutions.'' South Big Data Hub.
\url{https://southbigdatahub.org/resources/newsblog/keeping-data-science-broad-program}.

\leavevmode\hypertarget{ref-wickham2016r}{}%
Wickham, Hadley, and Garrett Grolemund. 2016. \emph{R for Data Science:
Import, Tidy, Transform, Visualize, and Model Data}. O'Reilly Media,
Inc.: Sebastopol, CA. \url{https://r4ds.had.co.nz}.

\leavevmode\hypertarget{ref-yan2019first}{}%
Yan, Donghui, and Gary E Davis. 2019. ``A First Course in Data
Science.'' \emph{Journal of Statistics Education} 27 (2): 99--109.
\url{https://doi.org/10.1080/10691898.2019.1623136}.
\end{cslreferences}

\bibliographystyle{apa}
\bibliography{references.bib}

\end{document}